\documentclass[preprint,12pt]{elsarticle}

\usepackage{epsfig}
\usepackage{amsmath}
\usepackage{graphicx}
\usepackage{hyperref}

\usepackage[lmargin=2cm,rmargin=2cm,tmargin=2.5cm,bmargin=3.5cm]{geometry}

\usepackage{epstopdf}
\journal{Physics Letters B}
\begin{document}

\begin{frontmatter}

\title{Nuclear suppression from coherent $J/\psi$ photoproduction at the Large Hadron Collider}

\author[label1]{V. Guzey}
\author[label1]{E. Kryshen}
\author[label2]{M. Strikman}
\author[label1]{M. Zhalov}

\address[label1]{National Research Center ``Kurchatov Institute'', Petersburg
 Nuclear Physics Institute (PNPI), Gatchina, 188300, Russia}
 
\address[label2]{Department of Physics, The Pennsylvania State University, State College, PA 16802, USA}

\begin{abstract}

Using the data on coherent $J/\psi$ photoproduction in Pb-Pb ultraperipheral collisions (UPCs) obtained in Runs 1 and 2 at the 
Large Hadron Collider (LHC), we
determined with a good accuracy the nuclear suppression factor of $S_{Pb}(x)$ in a wide range of the momentum fraction 
$x$, $10^{-5} \leq x \leq 0.04$. 
In the small-$x$ region $x < 10^{-3}$, our $\chi^2$ fit favors a flat form of $S_{Pb}(x) \approx 0.6$ with approximately a $5$\% accuracy 
for $x=6 \times 10^{-4} - 10^{-3} $ and a $25$\% error at $x=10^{-4}$. At the same time, uncertainties of the fit do not  
exclude a slow decrease of $S_{Pb}(x)$ in the small-$x$ limit.
 At large $x$, $S_{Pb}(x)$ is constrained to 
better than 10\% precision up to $x=0.04$ and is
also consistent with the value of $S_{Pb}(x)$ at $\langle x \rangle =0.042$, which we extract from the Fermilab data on the $A$ dependence 
of the cross section of coherent $J/\psi$ photoproduction on fixed nuclear targets. 
The resulting uncertainties on $S_{Pb}(x)$ are small, which indicates the potential of 
the LHC data on coherent charmonium photoproduction in Pb-Pb UPCs to provide additional constraints on small-$x$ nPDFs.
We explicitly demonstrate this using as an example the EPPS16 and nCTEQ16 nuclear parton distribution functions, whose uncertainties
decrease severalfold after the Bayesian reweighting of the discussed UPC data.

\end{abstract}

\begin{keyword}

ultraperipheral heavy-ion collisions \sep charmonium photoproduction \sep nuclear shadowing \sep parton distributions in nuclei

\end{keyword}

\end{frontmatter}

\section{Introduction}
\label{sec:intro}

Determination of nuclear parton distribution functions (nPDFs) is an important topic of phenomenology of high energy
nuclear physics. In the context of Quantum Chromodynamics (QCD), collinear nPDFs are universal quantities encoding the microscopic quark and gluon structure of nuclei probed in various hard processes. 
One usually determines nPDFs using so-called global QCD fits to available data~\cite{deFlorian:2003qf,Hirai:2007sx,Eskola:2009uj,deFlorian:2011fp,Kovarik:2015cma,Khanpour:2016pph,Eskola:2016oht,AbdulKhalek:2019mzd}.
However, because of limited kinematic coverage of the available data and largely indirect determination of the gluon distribution, 
nPDFs are currently known with large uncertainties. 
Recent QCD analyses of the proton-nucleus ($pA$) data collected during Runs 1 and 2 at the LHC~\cite{Eskola:2013aya,Helenius:2014qla,Armesto:2015lrg,Kusina:2016fxy,Kusina:2017gkz,Eskola:2019dui,Eskola:2019bgf,Kusina:2020lyz} showed 
that while they provide certain new restrictions on nPDFs, the remaining uncertainties are still significant.
Alternatively, 
 nuclear structure functions and nPDFs at small $x$ can be theoretically predicted using models of nuclear shadowing, which are based on its connection to diffraction~\cite{Frankfurt:1998ym,Piller:1999wx,Armesto:2003fi,Frankfurt:2011cs}. In particular, in Ref.~\cite{Frankfurt:2011cs}, the use of QCD factorization theorems allowed one to connect the leading-twist nuclear shadowing of nPDFs to proton diffractive PDFs and, hence, 
 predict small-$x$ nPDFs with a small uncertainty.
This topic will be further pursued after luminosity and energy upgrades of the LHC~\cite{Citron:2018lsq}. 

In the limit of very high energies, one often uses the framework of the color dipole model, 
which allows one to study the proximity of the dipole-target interaction to 
the new QCD regime characterized by saturation of the gluon density, for reviews, see, e.g.~\cite{Gelis:2010nm,Albacete:2014fwa}. Establishing the pattern and signs of the saturation using HERA and LHC data remains a challenge and an active field of 
research, see, e.g.~\cite{Mantysaari:2018nng,Mantysaari:2019nnt}.

In the future, it is expected that nPDFs and possible signs of an onset of the gluon saturation 
will be explored with high precision and in a broad kinematic range using such lepton-nucleus colliders as the Electron-Ion Collider in the USA~\cite{Boer:2011fh,Accardi:2012qut} and 
the Large Hadron-Electron Collider (LHeC)~\cite{AbelleiraFernandez:2012cc} and Future Circular Collider (FCC)~\cite{Abada:2019lih} at CERN. Meanwhile it is important to utilize all existing capabilities of the LHC to constrain nPDFs including those provided by
ultraperipheral collisions (UPCs) of heavy ions.

In UPCs, ions in colliding beams interact at large distances between their centers in the transverse plane (large impact parameters) so that strong hadron interactions are suppressed leading to the dominance of long-distance electromagnetic processes induced by ultrarelativistic nuclei, which in the equivalent photon approximation are characterized by fluxes of
quasireal photons of high intensity and energy. Thus, it gives an opportunity to study photon-nucleus scattering and nPDFs 
at unprecedentedly high energies~\cite{Baltz:2007kq}.
In particular, QCD analyses of photoproduction of heavy quarkonia~\cite{Frankfurt:2001db,Guzey:2013xba,Guzey:2013qza}
and inclusive  and diffractive dijet photoproduction~\cite{Guzey:2018dlm,Guzey:2019kik,Guzey:2016tek} at the LHC
provided new information on nuclear gluon and quark distributions at small $x$.

This work continues and extends our phenomenological studies of nuclear suppression in coherent $J/\psi$ photoproduction on
nuclei at the LHC~\cite{Guzey:2013xba,Guzey:2013qza} by including in the analysis all the data available to date on
the rapidity $y$ dependence of the cross section of coherent $J/\psi$ photoproduction in Pb-Pb UPCs at $\sqrt{s_{NN}}=2.76$ TeV~\cite{Abbas:2013oua,Abelev:2012ba,Khachatryan:2016qhq} and $\sqrt{s_{NN}}=5.02$ TeV~\cite{Acharya:2019vlb,LHCb:2018ofh,Acharya:2021ugn}.
As a cross-check and reference point at lower energies, we test our results against the data on the mass number $A$ dependence of the cross section of coherent $J/\psi$ photoproduction
on fixed nuclear targets (Be, Fe, and Pb) obtained at Fermilab~\cite{Sokoloff:1986bu}.
Note that it would also be very beneficial to collect high statistics on $J/\psi$ photoproduction in heavy-ion UPCs at RHIC
because it would cover the $x$ range of $x \sim 0.015$ at $y=0$ and help to further constrain the results of our analysis, 
provided the data is enough accurate.
Expressing our results in terms of the nuclear suppression factor of $S_{Pb}(x)$, we determine $S_{Pb}(x)$ with a good 
accuracy in a wide range of $x$, $10^{-5} \leq x \leq 0.04$. 
The resulting uncertainties are much smaller than those of nPDFs for these values of $x$, which indicates the potential of 
the LHC data on coherent charmonium photoproduction in Pb-Pb UPCs to provide new constraints on small-$x$ nPDFs and possible 
signs of saturation of the gluon density. We explicitly demonstrate this using the Bayesian reweighting of the discussed UPC data
and the EPPS16 and nCTEQ15 nPDFs, whose uncertainties become dramatically reduced.

\section{Nuclear suppression factor for coherent $J/\psi$ photoproduction on nuclei}
\label{sec:nucl_supp}

In UPCs, both colliding nuclei serve as a source of quasi-real photons and a target. Therefore, using the method of
equivalent photons~\cite{Budnev:1974de,Vidovic:1992ik},
 the cross section of
coherent $J/\psi$ photoproduction in symmetric Pb-Pb UPCs is given by a sum of the following two terms
\begin{equation}
\frac{d\sigma_{AA \to J/\psi AA}(\sqrt{s_{NN}},y)}{dy}=N_{\gamma/A}(W_{\gamma p}^{+}) \sigma_{\gamma A \to J/\psi A}(W_{\gamma p}^{+}) + N_{\gamma/A}(W_{\gamma p}^{-}) \sigma_{\gamma A \to J/\psi A}(W_{\gamma p}^{-}) \,,
\label{eq:sigma_AA}
\end{equation}
where $y$ is the rapidity of $J/\psi$, $N_{\gamma/A}(W_{\gamma p})$ is the photon flux, and $\sigma_{\gamma A \to J/\psi A}(W_{\gamma p})$ is the
photoproduction cross section containing all details of the strong photon-nucleus interaction and production of $J/\psi$.
Note that interference of the two terms in Eq.~(\ref{eq:sigma_AA}) is sizable only at very small values of the 
$J/\psi$ transverse momentum~\cite{Klein:1999gv} and hence can be safely neglected.

In the laboratory frame (coinciding with centre-of-mass system in our kinematics), the measured rapidity of $J/\psi$ can be related to the invariant photon-nucleon energy $W_{\gamma p}$,
\begin{equation}
W_{\gamma p}^{\pm}=\sqrt{2 E_N M_{J/\psi}}\,e^{\pm y/2} \,, 
\label{eq:W}
\end{equation}
where $E_N$ is the nuclear beam energy per nucleon and $M_{J/\psi}$ is the mass of $J/\psi$. The ambiguity in $W_{\gamma p}$  for $y \neq 0$ is a reflection of the presence of two terms in Eq.~(\ref{eq:sigma_AA}), where 
the first term corresponds to the right-moving photon source and the plus sign in Eq.~(\ref{eq:W}) and the second term corresponds to
the left-moving photon source and the minus sign in Eq.~(\ref{eq:W}) (provided that $y$ is defined with respect to the right-moving nucleus emitting the photon). 

To avoid inelastic strong ion-ion interactions destroying the coherence condition,
the photon flux in Eq.~(\ref{eq:sigma_AA}) is calculated as convolution over the impact parameter $\vec{b}$ of the flux of quasireal 
photons emitted by an ultrarelativistic charged ion $N_{\gamma/A}(\omega,\vec{b})$~\cite{Budnev:1974de,Vidovic:1992ik} with the probability 
not to have inelastic strong ion-ion interactions $\Gamma_{AA}(\vec{b})=\exp(-\sigma_{NN} \int d^2 \vec{b}_1 T_A(\vec{b}_1)
T_A(\vec{b}-\vec{b}_1))$:
\begin{equation}
N_{\gamma/A}(W_{\gamma p})=\int d^2 \vec{b}\, N_{\gamma/A}(\omega,\vec{b}) \Gamma_{AA}(\vec{b}) \,,
\label{eq:flux}
 \end{equation}
where $\omega=W_{\gamma p}^2/(4 E_N)$ is the photon energy; 
$\sigma_{NN}$ is the total nucleon-nucleon cross section; $T_A(\vec{b})=\int dz \rho_A(\vec{b},z)$ is the so-called nuclear optical density, which is calculated using the Woods-Saxon (two-parameter Fermi model) parametrization of the nuclear density 
$\rho_A$~\cite{DeJager:1987qc}.
One should emphasize that the precise determination of the photon flux using Eq.~(\ref{eq:flux}) in a wide range of $\omega$
is essential for the analysis of the present work. The validity of the equivalent photon approximation and a model~\cite{Pshenichnov:2001qd,Pshenichnov:2011zz} generalizing
Eq.~(\ref{eq:flux}) were successfully tested in
electromagnetic dissociation with neutron emission in Pb-Pb UPCs~\cite{ALICE:2012aa}.

The UPC cross section~(\ref{eq:sigma_AA}) is subject to nuclear modifications, which originate from the photon flux and 
the photoproduction cross section and which in general depend on the rapidity $y$ and the collision energy
$\sqrt{s_{NN}}$. To quantify the magnitude of nuclear corrections due to the strong dynamics encoded in the photoproduction cross section and to separate the two contributions in Eq.~(\ref{eq:sigma_AA}), it is convenient to introduce the
nuclear suppression factor of $S_{Pb}(x)$ by the following relation, see Refs.~\cite{Guzey:2013xba,Guzey:2013qza}:
\begin{equation}
S_{Pb}(x)=\sqrt{\frac{\sigma_{\gamma A \to J/\psi A}(W_{\gamma p})}{\sigma_{\gamma A \to J/\psi A}^{\rm IA}(W_{\gamma p})}} \,,
\label{eq:Sx}
\end{equation}
where $x=M_{J/\psi}^2/W_{\gamma p}^2$. The denominator in Eq.~(\ref{eq:Sx}) is the coherent $J/\psi$ photoproduction cross section 
in the impulse approximation (IA),
\begin{equation}
\sigma_{\gamma A \to J/\psi A}^{\rm IA}(W_{\gamma p})=\frac{d\sigma_{\gamma p \to J/\psi p}(W_{\gamma p},t=0)}{dt}
\int_{|t_{\rm min}|}^{\infty} dt |F_A(t)|^2 \,,
\label{eq:IA}
\end{equation}
where $F_A(t)$ is the nuclear elastic form factor and $|t_{\rm min}|=x^2 m_N^2$ is the minimal momentum transfer squared ($m_N$ is the nucleon mass). In our work, $F_A(t)$ was calculated using the Woods-Saxon parametrization of the nuclear density~\cite{DeJager:1987qc}.
The differential cross section of $J/\psi$ photoproduction on the proton was parametrized in the form~\cite{Guzey:2013xba}, which provides a good description of the available data at fixed targets~\cite{Camerini:1975cy,Binkley:1981kv,Aubert:1979ri} and at HERA~\cite{Chekanov:2002xi,Aktas:2005xu},
\begin{equation}
\frac{d\sigma_{\gamma p \to J/\psi p}(W_{\gamma p},t=0)}{dt}=C_0 \left[1.0-\frac{(M_{J/\psi}+m_N)^2}{W_{\gamma p}^2}\right]^{1.5}
 \left(W_{\gamma p}^2/W_0^2\right)^{\delta} \,,
 \label{eq:sigma_p}
\end{equation}
where $C_0=342 \pm 8$ nb/GeV$^2$, $\delta=0.40 \pm 0.01$, $W_0=100$ GeV.
For $W_{\gamma p} \leq 1$ TeV, this parametrization is consistent with a power-law fit to the $W$ dependence of the $\gamma p \to J/\psi p$ cross section
extracted from the LHCb data on coherent $J/\psi$ photoproduction in proton-proton UPCs at $\sqrt{s_{NN}}=7$ TeV~\cite{Aaij:2014iea} and $\sqrt{s_{NN}}=13$ TeV~\cite{Aaij:2018arx}. For higher photon energies $W_{\gamma p} > 1$ TeV, the extracted cross section 
shows a deviation from a pure power-law extrapolation of the HERA data, see the discussion in Ref.~\cite{Aaij:2018arx}. 
However, this region of $W_{\gamma p}$ is
not probed in the Pb-Pb UPC data and, hence, does not affect the results of our analysis.
Thus, the $\sigma_{\gamma A \to J/\psi A}^{\rm IA}(W_{\gamma p})$ cross section is evaluated 
%vg model-independently
using data-driven parameterizations of the nuclear form factor and the $\gamma p \to J/\psi p$ differential
cross section.

Introducing the UPC cross section in the impulse approximation $d\sigma_{AA \to J/\psi AA}^{\rm IA}/dy$, 
\begin{equation}
\frac{d\sigma_{AA \to J/\psi AA}^{\rm IA}(\sqrt{s_{NN}},y)}{dy}=N_{\gamma/A}(W_{\gamma p}^{+}) \sigma_{\gamma A \to J/\psi A}^{\rm IA}(W_{\gamma p}^{+}) + N_{\gamma/A}(W_{\gamma p}^{-}) \sigma_{\gamma A \to J/\psi A}^{\rm IA}(W_{\gamma p}^{-}) \,,
\label{eq:sigma_AA_IA}
\end{equation}
one can present the square root of the ratio of the UPCs cross sections entering Eqs.~(\ref{eq:sigma_AA}) and (\ref{eq:sigma_AA_IA}) 
in the following form
\begin{eqnarray}
&&\left(\frac{d\sigma_{AA \to J/\psi AA}(\sqrt{s_{NN}},y)/dy}{d\sigma_{AA \to J/\psi AA}^{\rm IA}(\sqrt{s_{NN}},y)/dy} \right)^{1/2} \nonumber\\
 &=&
\left(\frac{N_{\gamma/A}(W_{\gamma p}^{+}) S_{Pb}^2(x_{+}) \sigma_{\gamma A \to J/\psi A}^{\rm IA}(W_{\gamma p}^{+}) + N_{\gamma/A}(W_{\gamma p}^{-}) S_{Pb}^2(x_{-})\sigma_{\gamma A \to J/\psi A}^{\rm IA}(W_{\gamma p}^{-})}{N_{\gamma/A}(W_{\gamma p}^{+}) \sigma_{\gamma A \to J/\psi A}^{\rm IA}(W_{\gamma p}^{+}) + N_{\gamma/A}(W_{\gamma p}^{-}) \sigma_{\gamma A \to J/\psi A}^{\rm IA}(W_{\gamma p}^{-})} \right)^{1/2} \,,
\label{eq:R_Pb}
\end{eqnarray}
where $x_{\pm}=M_{J/\psi}^2/W_{\gamma p}^{\pm 2}$. Without loss of generality, we will use $y \geq 0$ and, hence,
$W_{\gamma p}^{+} \geq W_{\gamma p}^{-} $ and $x_{+} \leq x_{-}$.
The advantage of Eq.~(\ref{eq:R_Pb}) is that it relates the experimentally
measured UPC cross section ratio on the left-hand side to the nuclear suppression factor of $S_{Pb}(x)$ on the right-hand side.
However, it involves $S_{Pb}^2(x)$ at two different values of $x$ and is generally dominated by the $x_{-}$ contribution since 
$N_{\gamma/A}(W_{\gamma p}^{-}) \gg N_{\gamma/A}(W_{\gamma p}^{+})$, which complicates the separation of the $x_{+}$ and $x_{-}$ contributions and reliable extraction of the  $x_{+}$ term corresponding to higher energies.
Nevertheless, the use of all the available data on Pb-Pb UPCs collected during Runs 1 and 2 at the LHC along with a 
general parametrization of $S_{Pb}(x)$ allows us to extract $S_{Pb}(x)$ down to $x \approx 10^{-5}$ with a good  precision.
Note that the two contributions to the UPC cross section can also be separated by measuring ion-ion UPCs accompanied by mutual 
electromagnetic excitation of colliding ions followed by forward neutron emission~\cite{Guzey:2013jaa}. Unfortunately, the statistics of such
measurements is currently too low.

The UPC data used in our analysis include the ALICE~\cite{Abbas:2013oua,Abelev:2012ba} and CMS~\cite{Khachatryan:2016qhq} data at
$\sqrt{s_{NN}}=2.76$ TeV and the ALICE~\cite{Acharya:2019vlb,Acharya:2021ugn} and LHCb~\cite{LHCb:2018ofh} data at $\sqrt{s_{NN}}=5.02$ TeV.
It is summarized in Table~\ref{table:data} showing the rapidity intervals, the corresponding UPC cross sections $d\sigma_{AA \to J/\psi AA}/dy$, and the values
of $x_{+}$ and $x_{-}$. The last column gives the 
$[d\sigma_{AA \to J/\psi AA}/dy)/(d\sigma_{AA \to J/\psi AA}^{\rm IA}/dy)]^{1/2}$ cross section ratio 
calculated using Eq.~(\ref{eq:R_Pb}); the error is the sum of experimental statistical and systematic uncertainties
as well as the 5\% theoretical error on the IA cross section~\cite{Guzey:2013xba}
added in quadrature.
%vg 
Note that the latter error largely cancels in the ratio in the right-hand side of Eq.~(\ref{eq:R_Pb}), and, hence, does not lead to
additional uncertainties in our analysis. 

\begin{table}[t]
\caption{Summary of the data on the cross section of coherent $J/\psi$ photoproduction in Pb-Pb UPCs used in our analysis:
the rapidity intervals, the corresponding UPC cross sections $d\sigma_{AA \to J/\psi AA}/dy$, and the values
of $x_{+}$ and $x_{-}$. The last column is the
$[d\sigma_{AA \to J/\psi AA}/dy)/(d\sigma_{AA \to J/\psi AA}^{\rm IA}/dy)]^{1/2}$ cross section ratio 
calculated using Eq.~(\ref{eq:R_Pb}).}
\begin{small}
\begin{center}
\begin{tabular}{|c|c|c|c|c|}
\hline
Rapidity interval & $d\sigma/dy$, mb & Refs. & $(x_+,x_{-})$ & $\sqrt{(d\sigma/dy)/(d\sigma^{\rm IA}/dy)]}$ \\
%$[d\sigma/dy)/(d\sigma^{\rm IA}/dy)]^{1/2}$ \\
\hline
$-0.9 < y < 0.9$     & $2.38^{+0.34}_{-0.24} \, ({\rm stat+syst})$ & \cite{Abbas:2013oua} &
$(1.12 \times 10^{-3},1.12 \times 10^{-3})$ & $0.62 \pm 0.057$ \\
$ 1.8 < |y| < 2.3$   & $1.82 \pm 0.22 \, (\rm stat) \pm 0.20 \, (\rm syst)$ &  & $(1.44 \times 10^{-4}, 8.72 \times 10^{-3})$ & $0.69 \pm 0.069$ \\
& $\pm 0.19 \,(\rm theo)$ & \cite{Khachatryan:2016qhq} & & \\
$-3.6 < y < -2.6$    & $1.00 \pm 0.18 \, ({\rm stat})^{+0.24}_{-0.26} \, ({\rm syst})$ & \cite{Abelev:2012ba} & $(5.05 \times 10^{-5}, 2.49 \times 10^{-2})$ & $0.72 \pm 0.12$\\ 
\hline
\hline
$-4.00 < y < -3.75$  & $1.615 \pm 0.060 \, (\rm stat)^{+0.135}_{-0.147} \,(syst)$ & \cite{Acharya:2019vlb} & $(1.28 \times 10^{-5}, 2.97 \times 10^{-2})$ & $0.89 \pm 0.049$ \\
$-3.75 < y < -3.50$  & $1.938 \pm 0.042 \, (\rm stat)^{+0.166}_{-0.190} \, (syst)$ & & $(1.64 \times 10^{-5}, 2.31 \times 10^{-2})$ & $0.87 \pm  0.048$\\
$-3.50 < y < -3.25$  & $2.377 \pm 0.040 \, (\rm stat)^{+0.212}_{-0.229} \, (syst)$ & & $(2.11 \times 10^{-5}, 1.80 \times 10^{-2})$ & $0.87 \pm    0.048$ \\
$-3.25 < y < -3.00$  & $2.831 \pm 0.047 \, (\rm stat)^{+0.253}_{-0.280} \, (syst)$ & & $(2.71 \times 10^{-5}, 1.40 \times 10^{-2})$ & $0.86 \pm    0.048$\\
$-3.00 < y < -2.75$  & $3.018 \pm 0.061 \, (\rm stat)^{+0.259}_{-0.294} \, (syst)$ & &  $(3.48 \times 10^{-5}, 1.09 \times 10^{-2})$ & $0.81 \pm    0.045$\\
$-2.75 < y < -2.50$  & $3.531 \pm 0.139 \, (\rm stat)^{+0.294}_{-0.362} \, (syst)$ & & $(4.47 \times 10^{-5}, 0.85 \times 10^{-2})$ & $0.81 \pm    0.049$ \\
\hline
$ 2.00 < y < 2.50$ & $3.0 \pm 0.4\, (\rm stat) \pm 0.3\,(syst)$ & \cite{LHCb:2018ofh} &  $(6.50 \times 10^{-5}, 0.59 \times 10^{-2})$ & $0.67 \pm    0.059$\\
$ 2.50 < y < 3.00$ & $2.60 \pm 0.19\, (\rm stat) \pm 0.25\,(syst)$ & &  $(3.94 \times 10^{-5}, 0.96 \times 10^{-2})$
 & $0.72 \pm 0.047$\\
$ 3.00 < y < 3.50$ & $2.28 \pm 0.15\, (\rm stat) \pm 0.21\,(syst)$ & & $(2.39 \times 10^{-5}, 1.59 \times 10^{-2})$
 & $0.81 \pm 0.050$ \\
$ 3.50 < y < 4.00$ & $1.73 \pm 0.15\, (\rm stat) \pm 0.17\,(syst)$ & & $(1.45 \times 10^{-5}, 2.62 \times 10^{-2})$
& $0.87 \pm 0.062$ \\
$ 4.00 < y < 4.50$ & $1.10 \pm 0.22\, (\rm stat) \pm 0.13\,(syst)$ & & $(0.88 \times 10^{-5}, 4.32 \times 10^{-2})$ 
& $0.91 \pm 0.11$\\
\hline
$-0.15 < y < 0.15$ &  $4.075 \pm 0.114\, (\rm stat) \pm 0.231\,(syst)$ & \cite{Acharya:2021ugn} & $(6.17 \times 10^{-4}, 6.17 \times 10^{-4})$ & $0.63 \pm    0.025$\\
$0.15 < |y| < 0.35$ &  $4.211 \pm 0.114\, (\rm stat) \pm 0.239\,(syst)$ & & $(7.92 \times 10^{-4}, 4.80 \times 10^{-4})$ & $0.64 \pm    0.026$\\
$0.35 < |y| < 0.8$ &  $3.870 \pm 0.110\, (\rm stat) \pm 0.220\,(syst)$ &  & $(1.10 \times 10^{-3}, 3.47 \times 10^{-4})$ & $0.62 \pm    0.025$\\
\hline
\end{tabular}
\end{center}
\end{small}
\label{table:data}
\end{table}%

To constrain the nuclear suppression factor of $S_{Pb}(x)$ in as a broad range of $x$ as possible, we 
apply Eq.~(\ref{eq:R_Pb}) to the available UPC data listed in Table~\ref{table:data}.
Note that as explained in Refs.~\cite{Guzey:2013xba,Guzey:2013qza}, the ALICE data points at $y=0$ from Run 1~\cite{Abbas:2013oua}
and Run 2~\cite{Acharya:2021ugn} (the first and 15th entries in Table~\ref{table:data}, respectively)
 unambiguously and model-independently correspond to 
\begin{eqnarray}
S_{Pb}(x=0.00112)=0.62 \pm 0.057\,, \nonumber\\
S_{Pb}(x=6.17 \times 10^{-4})=0.63 \pm 0.025 \,.
\label{eq:S_Pb_2}
\end{eqnarray}

The shape of $S_{Pb}(x)$ as a function of $x$ is unconstrained. In this work, we use the following 
simple piece-wise parametrization of $S_{Pb}(x)$
\begin{equation}
S_{Pb}(x) = \left\{
\begin{array}{ll} 
 a+b_1 \ln(x_1/x_0)+b_2 \ln (x/x_1)\,, & {\rm for} \ x \geq x_1  \\
 a+b_1 \ln(x/x_0) \,, & {\rm for} \ x_1 > x >  x_0 \,, \\
 a+c \ln(x/x_0) \,, & {\rm for} \ x \leq  x_0 \,,
 \end{array}
\right.
\label{eq:Sx_fit} 
\end{equation}
where $x_0 \approx 5 \times 10^{-4}-10^{-3}$, $x_1 \approx 0.01-0.05$,  and $c \geq 0$.
 Our fit function contains four free parameters: $a$ is determined
by Eq.~(\ref{eq:S_Pb_2}) and $b_{1,2}$ and $c$ are largely constrained by the low-energy $W_{\gamma p}^{-}$ and the high-energy $W_{\gamma p}^{+}$ contributions to the UPC cross sections, respectively. 
The ranges of tried values of $x_0$ and $x_1$ are motivated by Eq.~(\ref{eq:S_Pb_2}) and the shapes of the  EPPS16~\cite{Eskola:2016oht} and nCTEQ15~\cite{Kovarik:2015cma} nPDFs, respectively.
The form of the parametrization in  Eq.~(\ref{eq:Sx_fit}) 
also allows for different slopes of $S_{Pb}(x)$ in different regions of $x$, including the possibility of a slow $x$ dependence
for $x \leq  x_0$ and a rapid $x$ dependence for $x \geq x_1$, which is analogous to those encoded in the 
EPPS16 and nCTEQ15 nPDFs.
 Note that we require that $c \geq 0$, which is part of our model assuming 
that the factor of nuclear suppression is a monotonic function of $x$ in the considered interval of $10^{-5} < x < 0.05$.

The results of the fit are presented in Tables \ref{table:Sx} and \ref{table:chi2}, where Table \ref{table:Sx} summarizes 
the values of the parameters of the fit and Table \ref{table:chi2} gives the values of $\chi^2$ for 17 data points used in the fit.
Several comments are in order here. First, while we chose to use $x_0=0.00112$ and $x_1=0.01$, the values of the parameters
$a$, $b_1$, $b_2$, and $c$ and $\chi^2$ weakly depend on the exact values of $x_0$ and $x_1$ as long as they are varied in the intervals 
$x_0 \approx 5 \times 10^{-4}-10^{-3}$ and $x_1 \approx 0.01-0.05$. Second, while the fit returns sufficiently good 
$\chi^2/{\rm d.o.f.}=0.76$,
more than half of the total value of $\chi^2$ comes from
the Run 2 ALICE data at forward rapidity. This indicates that the Run 2 ALICE and LHCb data at forward rapidity are somewhat mutually
inconsistent, see also the lower panel of  Fig.~\ref{fig:Sigma_pb208_2020}. For instance, decreasing the normalization of these 
ALICE data points by the typical value of the correlated
systematic uncertainty of  6\% will reduce by half the value of $\chi^2$ without noticeably changing the values of the fit parameters.

\begin{table}[t]
\caption{Parameters of the fit in Eq.~(\ref{eq:Sx_fit}).}
\begin{center}
\begin{tabular}{|c|c|}
\hline
Fit parameters & \\
\hline
$x_0$    &   $0.00112$ \\
$x_1$    &   $0.01$ \\ 
$a$      &   $0.63 \pm 0.014$ \\
$b_1$    &   $0.097 \pm 0.017$ \\
$b_2$    &   $0.053 \pm 0.045$ \\
$c$      &   $0 \pm 0.066$ \\
\hline
\end{tabular}
\end{center}
\label{table:Sx}
\end{table}%

\begin{table}[h]
\caption{The values of $\chi^2$ for individual fitted data points and the total $\chi^2$ for the fit in Eq.~(\ref{eq:Sx_fit}) and Table~\ref{table:Sx}. }
\begin{center}
\begin{tabular}{|c|c|}
\hline
Data point & $\chi^2$  \\
\hline
1 & $5.9 \times 10^{-2}$ \\
2 & 1.4 \\
3 & 1.7 \\
\hline
4 & 2.7  \\
5 & 1.2  \\
6 & 0.42 \\
7 & 1.1  \\
8 & 0.32 \\
9 & 1.0 \\
\hline 
10 & 0.75 \\
11 & 1.2 \\
12 & $8.3 \times 10^{-2}$ \\
13 & $2.2 \times 10^{-2}$ \\
14 & $4.3 \times 10^{-3}$ \\
\hline
15 & $2.3 \times 10^{-3}$ \\
16 & 0.26 \\
17 & $9.8 \times 10^{-2}$ \\
\hline
Total & 8.4 \\
\hline
\end{tabular}
\end{center}
\label{table:chi2}
\end{table}%

\begin{figure}[t]
\begin{center}
\epsfig{file=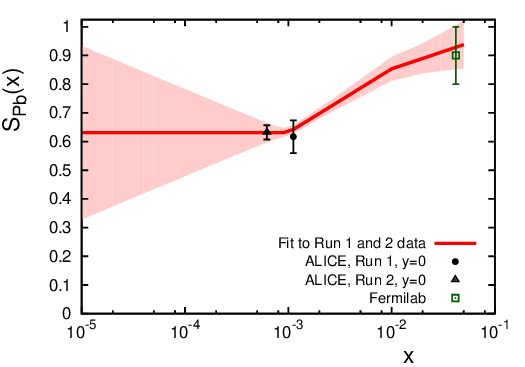,scale=1.4}
\caption{$S_{Pb}(x)$ as a function of $x$, see Eq.~(\ref{eq:Sx_fit}), fitted to all available data on 
coherent $J/\psi$ photoproduction in Pb-Pb UPCs at the LHC, see text for details. The shaded band represents the
uncertainty due to errors of the fit parameters.
The ALICE data points at $y=0$ from Run 1 and Run 2 are shown by the filled circle and an open triangle with the associated error,
respectively,
 see Eq.~(\ref{eq:S_Pb_2}).
The Fermilab data converted into $S_{Pb}(x)$ at $\langle x \rangle =0.042$ is shown by an open square with the corresponding uncertainty,
see Eq.~(\ref{eq:S_Pb_Fermilab}). 
}
\label{fig:Fit_final}
\end{center}
\end{figure}

The resulting $S_{Pb}(x)$ as a function of $x$ is shown in Fig.~\ref{fig:Fit_final};
the shaded band represents the uncertainty of the fit. Also, we show the values of $S_{Pb}(x=0.00112)$ and 
$S_{Pb}(x=6.17 \times 10^{-4})$ extracted from the Run 1 and Run 2 ALICE data at $y=0$, see Eq.~(\ref{eq:S_Pb_2}),
 and $S_{Pb}(x=0.042)$ 
determined using the fixed-target Fermilab data [open square with the corresponding uncertainty, see Eq.~(\ref{eq:S_Pb_Fermilab}) and the discussion below].

Several features of the obtained results are noteworthy. First, the fit favors a flat behavior of
$S_{Pb}(x) \approx 0.6$ for $x \leq 0.001$. 
However, within significant uncertainties at small $x$, 
a slow decrease of $S_{Pb}(x)$ in the small-$x$ limit is also not excluded.
This  agrees with the small-$x$ behavior of the $g_A(x,\mu^2)/[A g_p(x,\mu^2)]$ ratio of the nuclear and proton 
gluon distributions assumed in the EPS09~\cite{Eskola:2009uj}, EPPS16~\cite{Eskola:2016oht}, and nCTEQ15~\cite{Kovarik:2015cma} nPDFs
and is consistent within uncertainties with predictions of the leading twist model of nuclear shadowing~\cite{Frankfurt:2011cs};
see also the discussion in Sec.~\ref{sec:npdfs}.
Second, the use of the Run 1 and 2 data allowed us to determine $S_{Pb}(x)$ with a reasonable precision
in a wide range of $x$.
In particular, in the small-$x$ region $x < 10^{-3}$, $S_{Pb}(x)$ is determined with approximately a $5$\% accuracy 
at $x=6 \times 10^{-4} - 10^{-3}$ and a $25$\% accuracy at $x=10^{-4}$. At large $x$, $S_{Pb}(x)$ is constrained to 
better than 10\% precision up to $x=0.04$.

Figure~\ref{fig:Sigma_pb208_2020} demonstrates how well the calculation of the cross section of coherent $J/\psi$ photoproduction in Pb-Pb UPCs using Eq.~(\ref{eq:sigma_AA}) with the nuclear suppression factor of $S_{Pb}(x)$ describes the available
Run 1 (upper panel) and Run 2 (lower panel) LHC data.
It shows $d\sigma_{AA \to J/\psi AA}(\sqrt{s_{NN}},y)/dy$ as a function of $|y|$, where the solid lines 
and the shaded band correspond to the fit for $S_{Pb}(x)$ and its uncertainty presented in Fig.~\ref{fig:Fit_final}.
One can see from the figure that within the experimental and fit uncertainties, one obtains a good description of the data.
An examination of the lower panel of Fig.~\ref{fig:Sigma_pb208_2020} demonstrates that the Run 2 ALICE data points lie systematically higher than the LHCb points (see the discussion above).

\begin{figure}[t]
\begin{center}
\epsfig{file=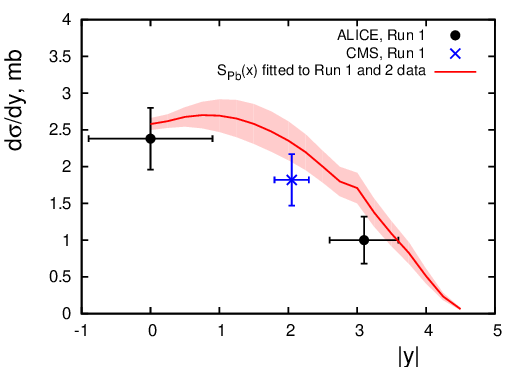,scale=1.4}
\epsfig{file=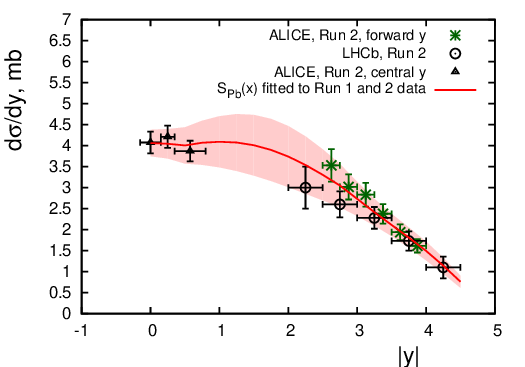,scale=1.4}
\caption{The $d\sigma_{AA \to J/\psi AA}(\sqrt{s_{NN}},y)/dy$ cross section of coherent $J/\psi$ photoproduction in Pb-Pb UPCs
as a function of $|y|$: the calculation using Eq.~(\ref{eq:sigma_AA}) with the nuclear suppression factor of $S_{Pb}(x)$ vs.~the Run 1 (upper panel) and Run 2 LHC data (lower panel). The shaded band shows the  uncertainty in the UPC cross section due to
the uncertainty of the fit, see Table~\ref{table:Sx} and Fig.~\ref{fig:Fit_final}.}
\label{fig:Sigma_pb208_2020}
\end{center}
\end{figure}

In addition to the UPC data in Table~\ref{table:data}, $S_{Pb}(x)$ at large $x$ can be further constrained
 using the Fermilab data on coherent
$J/\psi$ photoproduction by a photon beam with the average energy of 120 GeV on fixed nuclear targets of beryllium (Be), iron (Fe), 
and lead (Pb)~\cite{Sokoloff:1986bu}, where the corresponding average value of $x$ is $\langle x \rangle =0.042$.
The measured yields are normalized to the incoherent cross section on Be and their nuclear 
mass number $A$ dependence is fitted to the power law $A^{\alpha}$ with $\alpha=1.40 \pm 0.06 \pm 0.04$.
It is close to the expectation of the impulse approximation: a $\chi^2$ fit to the $A$ dependence given by 
Eq.~(\ref{eq:IA}) gives $\sigma_{\gamma A \to J/\psi A}^{\rm IA} \propto A^{1.44}$.
This indicates that nuclear corrections at these values of $W_{\gamma p}$ and $x$ are small.

To convert this result into the value of $S_{Pb}(x)$ at $\langle x \rangle =0.042$, we calculate it using the optical limit of the Glauber model,
see, e.g., Ref.~\cite{Guzey:2013xba},
\begin{equation}
S_{A}(x)=\frac{2 \int d^2 \vec{b}\, (1-e^{-\frac{\sigma_{VN}(W_{\gamma p})}{2} T_A(b)})}{A\sigma_{VN}(W_{\gamma p})} \,,
\label{eq:Glauber}
\end{equation} 
where $\sigma_{VN}(W_{\gamma p})$ is the charmonium-nucleon cross section, which we keep as a free parameter.
While the application of such an approach to charmonium photoproduction at high energies is questionable, 
it provides an adequate estimate of $\sigma_{J/\psi N}$ at the considered medium energy.  
It is known very well that the effect of nuclear shadowing encoded in Eq.~(\ref{eq:Glauber}) slows down the $A$ dependence
of hadron-nucleus cross sections: the stronger the nuclear absorption (the larger the value of $\sigma_{VN}$), the slower the $A$
dependence of $S_{A}(x)$. 
In our analysis, we varied the value of $\sigma_{VN}$ so that the $A$ dependence of 
the product $\sigma_{\gamma A \to J/\psi A}^{\rm IA} S_A^2(x)$
reproduces that of the Fermilab data and found that $\sigma_{VN}(\langle W_{\gamma p}\rangle =16.4 \ {\rm GeV})=3 \pm 3$ mb. 
Note that this value agrees with the charmonium-nucleon cross section $\sigma_{J/\psi N}$ obtained in the generalized vector dominance model and the 
coupled-channel generalized Glauber model framework~\cite{Frankfurt:2003uj}
and in the QCD dipole formalism~\cite{Kopeliovich:2020has}.
Substituting it in Eq.~(\ref{eq:Glauber}), we 
find that
\begin{equation}
S_{Pb}(\langle x \rangle =0.042)=0.90 \pm 0.10 \,.
\label{eq:S_Pb_Fermilab}
\end{equation}
As can be seen in Fig.~\ref{fig:Fit_final}, this value is consistent with the result of the fit to the LHC UPC data.

It is important to mention the analysis of~\cite{Contreras:2016pkc}, where the nuclear suppression factor of $S_{Pb}(x)$ was extracted from measurements of
coherent $J/\psi$ photoproduction in ultraperipheral and peripheral Pb-Pb collisions at the LHC at $\sqrt{s_{NN}}=2.76$ TeV.
It was found that $S_{Pb}(x=0.029)=0.74 \pm 0.07$, $S_{Pb}(x=0.0011)=0.62 \pm 0.04$,
 and $S_{Pb}(x=4.4 \times 10^{-5})=0.48 \pm 0.10$. Within uncertainties, our results at $x=0.029$ and $x=4.4 \times 10^{-5}$
 are consistent; at $x=0.0011$, the present analysis and that of~\cite{Contreras:2016pkc} reproduced the finding of~\cite{Guzey:2013xba}.

At the same time, the dipole model generally predicts a somewhat smaller nuclear 
suppression~\cite{Lappi:2013am,Goncalves:2014wna,Luszczak:2019vdc,Sambasivam:2019gdd}, whose
magnitude significantly depends on details of the model implementation including the choice of
the charmonium wave function and the dipole cross section.

\section{Implications for nuclear PDFs}
\label{sec:npdfs}

As discussed in the Introduction, coherent $J/\psi$ photoproduction in Pb-Pb UPCs at the LHC can be used to obtain new constraints
on the nuclear gluon distribution at small $x$. Indeed, in the leading logarithmic approximation of perturbative pQCD and 
in the static limit for the charmonium wave function, the cross section of exclusive $J/\psi$ photoproduction is proportional 
to the gluon density of the target squared~\cite{Ryskin:1992ui}
\begin{equation}
\frac{d\sigma_{\gamma T \to J/\psi T}(W_{\gamma p},t=0)}{dt} \propto [g_T(x,\mu^2)]^2 \,,
\label{eq:rel1}
\end{equation}
where $x=M_{J/\psi}^2/W_{\gamma p}^2$; $\mu$ is the factorization (resolution) scale determined by the mass of the charm quark.
Applying this result to nuclear targets and accounting for the transverse momentum dependence via the nuclear form factor
$F_A(t)$, one obtains~\cite{Guzey:2013qza}
\begin{equation}
\sigma_{\gamma A \to J/\psi A}(W_{\gamma p}) = \kappa_{A/N}^2 \frac{d\sigma_{\gamma p \to J/\psi p}(W_{\gamma p},t=0)}{dt}
\left[\frac{g_A(x,\mu^2)}{A g_N(x,\mu^2)}\right]^2 \int_{|t_{\rm min}|}^{\infty} dt  |F_A(t)|^2 \,,
\label{eq:rel2}
\end{equation}
where $\kappa_{A/N}^2=(1+\eta_A^2) {\bar R}_{g,A}^2/[(1+\eta_p^2) {\bar R}_{g,p}^2]$ is a factor taking into account the slightly different $x$ dependence of the nuclear and proton gluon distributions ($\eta$ is the ratio of the real to the imaginary parts of the
$\gamma T \to J/\psi T$ amplitude, ${\bar R}_{g}$ is a phenomenological enhancement factor relating the usual gluon density to the gluon generalized parton distribution~\cite{Shuvaev:1999ce}).
In the case, when the gluon distributions in a nucleus and the proton have a similar small-$x$ behavior, i.e., 
$R_g(x,\mu^2)=g_A(x,\mu^2)/[A g_N(x,\mu^2)]$ is a flat function of $x$ (as in the case of the EPPS16 and nCTEQ15 nPDFs), $\kappa_{A/N} =1$ with a good precision.
%vg
Indeed, both $\eta$ and ${\bar R}_{g}$ are determined by the $W_{\gamma p}$ dependence of the $\gamma T \to J/\psi T$ amplitude, which in turn is determined by the $x$ dependence of the gluon density. If this dependence is the same for the nucleus and proton 
targets, which happens when $R_g(x,\mu^2)$ is a flat function of $x$, then $\eta_A=\eta_p$ and ${\bar R}_{g,A}={\bar R}_{g,p}$
 and, hence, $\kappa_{A/N}=1$. In the case of the leading twist model, nuclear shadowing somewhat slows down the $x$ dependence
 of the $\gamma T \to J/\psi T$ amplitude on the nucleus compared to that on the proton, which leads to 
 $\eta_A < \eta_p$ and ${\bar R}_{g,A}< {\bar R}_{g,p}$ resulting in $\kappa_{A/N} < 1$.

The relations of Eqs.~(\ref{eq:rel1}) and (\ref{eq:rel2}) are subject to several types of corrections including
next-to-leading order (NLO) QCD corrections~\cite{Ivanov:2004vd,Jones:2015nna}, a model-dependent relation between generalized parton distributions (GPDs) and usual parton distributions, and  relativistic corrections to the charmonium wave function~\cite{Hoodbhoy:1996zg,Frankfurt:1995jw,Frankfurt:1997fj,Krelina:2019egg,Lappi:2020ufv}. To minimize their effect, 
it is advantageous to consider the nuclear suppression factor of $S_{Pb}(x)$~(\ref{eq:Sx}), where it is expected that almost all kinematic factors and corrections cancel in the ratio of the nuclear and proton cross sections. This establishes a direct correspondence between the suppression factor of $S_{Pb}(x)$ and the ratio of the nuclear and nucleon gluon distributions $R_g(x,\mu^2)=g_A(x,\mu^2)/[A g_N(x,\mu^2)]$~\cite{Guzey:2013xba}. Indeed, using Eqs.~(\ref{eq:Sx}), (\ref{eq:IA}), and (\ref{eq:rel2}), one readily obtains 
\begin{equation}
S_{Pb}(x)=\kappa_{A/N} R_g(x,\mu^2) \,.
\label{eq:rel3}
\end{equation}
Following the analysis of Ref.~\cite{Guzey:2013qza}, we take advantage of ambiguity in the exact values of the scale $\mu$ and
take $\mu^2=3$ GeV$^2$ to best reproduce the available HERA and LHCb data on the $W_{\gamma p}$ dependence of the cross section
of exclusive $J/\psi$ photoproduction on the proton.

\begin{figure}[t]
\begin{center}
\epsfig{file=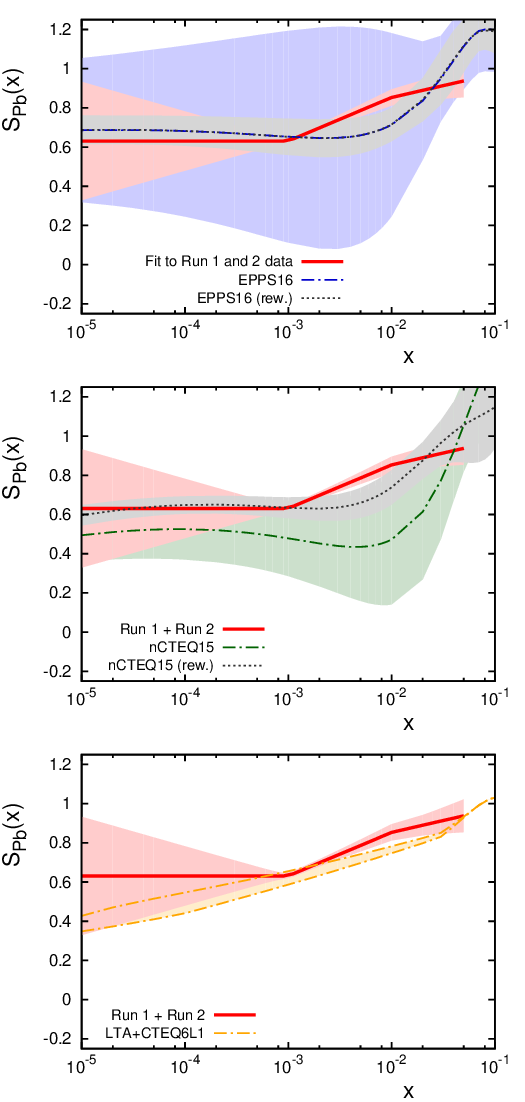,scale=1.05}
\caption{$S_{Pb}(x)$ and the $R_g(x,\mu^2)=g_A(x,\mu^2)/[A g_N(x,\mu^2)]$ ratios of the nuclear and nucleon gluon distributions
 as functions of $x$, which were evaluated using the EPPS16 (top) and nCTEQ15 (middle) nPDFs, and predictions of the leading twist model of nuclear shadowing (bottom) at $\mu^2=3$ GeV$^2$. In the upper and middle panels, 
 the dot-dashed curves and the outer shaded bands give 
  the central values and uncertainties of the corresponding nPDFs, respectively; 
  the dotted curves and the inner bands show the result of the reweighting, see text for details. 
}
\label{fig:Fit_comp}
\end{center}
\end{figure}

Figure~\ref{fig:Fit_comp} compares the results of our extraction of $S_{Pb}(x)$ with those of Eq.~(\ref{eq:rel3}). 
For the latter, for $R_g(x,\mu^2)$ we used the EPPS16~\cite{Eskola:2016oht} and nCTEQ15~\cite{Kovarik:2015cma}
nPDFs and predictions of the leading twist model of nuclear shadowing~\cite{Frankfurt:2011cs}. 
In the upper and middle panels, the dot-dashed curves give the central values of the corresponding nPDFs and the outer shaded bands 
represent their uncertainty, which is quite significant.  Note that as we discussed above, $\kappa_{A/N} =1$ 
for EPPS16 and nCTEQ15 and $\kappa_{A/N} =0.9$~\cite{Guzey:2013qza} for 
the leading twist nuclear shadowing model.

Several features of the presented results deserve to be pointed out.
First, in the very small $x$ region of $x < 10^{-3}$, the shapes of $S_{Pb}(x)$ and $R_g(x,\mu^2)$ for EPPS16 and nCTEQ15 nPDFs are 
similar predicting a flat form of the $x$ dependence; moreover, in the case of EPPS16, the magnitudes of
$S_{Pb}(x)$ and $R_g(x,\mu^2)$ are also very close.
Second, while the shapes of the EPPS16 and nCTEQ15 $R_g(x,\mu^2)$ are similar, 
the latter corresponds to the stronger nuclear shadowing (suppression) at $\mu^2=3$ GeV$^2$.
 Third, while predictions of the leading twist model agree with $S_{Pb}(x)$ for  $x \geq 10^{-3}$, the former predicts 
 $R_g(x,\mu^2)$, which noticeably decreases as $x$ is decreased (the two are still consistent within the significant uncertainty 
 of $S_{Pb}(x)$).

One can see from Fig.~\ref{fig:Fit_comp} that  the uncertainty of $R_g(x,\mu^2)$ for $x \geq 10^{-4}$ is much larger than that of $S_{Pb}(x)$, which indicates that the UPC data on coherent $J/\psi$ photoproduction in Pb-Pb UPCs at the LHC
 can potentially significantly reduce the current large uncertainty of the nuclear gluon distribution.
 To demonstrate this, we apply the standard Bayesian reweighting technique, see, e.g., Refs.~\cite{Kusina:2016fxy,Guzey:2019kik,Armesto:2013kqa,Paukkunen:2014zia}, to the discussed UPC data.
  Below we outline main steps of the method.
  
For a given set of nPDFs, one generates a large number of replicas $N_{\rm rep}$ (one usually takes $N_{\rm rep}=10,000$), which are
labeled by index $k$,
\begin{equation}
g_A^k(x,\mu^2) =g_A^0(x,\mu^2)+\frac{1}{2}\sum_{i=1}^{N} \left(g_A^{i+}(x,\mu^2)-g_A^{i-}(x,\mu^2)\right) R_{ki} \,,
\label{eq:rew1} 
\end{equation}
where $g_A^0(x,\mu^2)$ and $g_A^{i\pm}(x,\mu^2)$ are the central value and error PDFs corresponding to the eigenvector $i$
[the number of eigenvectors (fit parameters) is $N=20$ for EPPS16 and $N=16$ for nCTEQ15];
$R_{ki}$ are random numbers from the normal distribution centered at zero with the standard deviation of unity.
For each replica, we estimate how well it reproduces the ratios $\sqrt{(d\sigma/dy)/(d\sigma^{\rm IA}/dy)]}$ in Table~\ref{table:data}
 by calculating the corresponding $\chi_k^2$
\begin{equation}
\chi^2_k=\sum_{j=1}^{N_{\rm data}}\frac{
\left(\sqrt{(d\sigma/dy)/(d\sigma^{\rm IA}/dy)}^{(j)}-R_{Pb,k}^{(j)}\right)^2}
{\left(\delta\sqrt{(d\sigma/dy)/(d\sigma^{\rm IA}/dy)}^{(j)}\right)^2} \,,
 \label{eq:rew2} 
\end{equation}
where $j$ labels the data  points ($N_{\rm data}=17$); $\sqrt{(d\sigma/dy)/(d\sigma^{\rm IA}/dy)}^{(j)}$ and
$\delta\sqrt{(d\sigma/dy)/(d\sigma^{\rm IA}/dy)}^{(j)}$ are the cross section ratios and their uncertainties given by the last
column in Table~\ref{table:data};
$R_{Pb,k}^{(j)}$ stand for the right-hand side of Eq.~(\ref{eq:R_Pb}) evaluated using $S_{Pb}(x)=g_A^k(x,\mu^2)/[A g_N(x,\mu^2)]$ 
at point $j$ [see Eq.~(\ref{eq:rel1})].
Based on these $\chi_k^2$, one assigns each replica its statistical weight $w_k$, 
\begin{equation}
w_k=N_{\rm norm} e^{-\frac{1}{2} \chi^2_k/T} \,,
\label{eq:rew3} 
\end{equation}
where $T$ is the tolerance associated with a given set of PDFs, in particular, 
$T=52$ for EPPS16 and $T=35$ for nCTEQ15;
$N_{\rm norm}=N_{\rm rep} (\sum_i e^{-\frac{1}{2} \chi^2_i/T})^{-1}$
is the normalization constant chosen to satisfy the condition $\sum_k w_k=N_{\rm rep}$. 

The essence of the reweighting method is that instead of performing a new global QCD fit of nPDFs, 
 one can quantify the influence of the UPC data, 
which were not used in the original fits, on the EPPS16 and nCTEQ15 nPDFs.
 Using the weights $w_k$, one calculates the new, reweighted central values and uncertainties of the nuclear gluon distributions 
 \begin{eqnarray}
\langle g_A(x,\mu^2) \rangle &=&\frac{1}{N_{\rm rep}} \sum_{k=1}^{N_{\rm rep}} w_k g_A^k(x,\mu^2) \,, \nonumber \\
\delta \langle g_A(x,\mu^2) \rangle &=& \left[\frac{1}{N_{\rm rep}} \sum_{k=1}^{N_{\rm rep}} w_k \left(g_A^k(x,\mu^2)-\langle g_A(x,\mu^2) \rangle \right)^2 \right]^{1/2}\,. \nonumber \\
\label{eq:rew4}
 \end{eqnarray}

The results of this reweighting procedure are shown in the upper (EPPS16) and middle (nCTEQ15) panels of Fig.~\ref{fig:Fit_comp} and represented by the gray dotted curves and the inner shaded error bands. One immediately sees from the figure that the inclusion 
of the discussed UPC data using the reweighting method leads to a severalfold reduction of the current large uncertainties of the EPPS16 and nCTEQ15 nPDFs. Also, while in the EPPS16 case the original and reweighted central values of the gluon distribution coincide, 
in the nCTEQ15 case the UPC data prefers somewhat larger values of the gluon distribution for small $x$
(this can also be deduced from a comparison to the fit function $S_{Pb}(x)$ in the middle panel of Fig.~\ref{fig:Fit_comp}).

\section{Conclusions}
\label{sec:conclusions}

In this work, we analyzed the Run 1 and 2 LHC data on coherent $J/\psi$ photoproduction in Pb-Pb UPCs in terms of the nuclear suppression factor of $S_{Pb}(x)$ using its generic parametrization and a $\chi^2$ fit to the ratios of the measured UPCs cross sections to those calculated in the impulse approximation.
It allowed us to determine $S_{Pb}(x)$ with a reasonable accuracy in a wide range of $x$, $10^{-5} \leq x \leq 0.04$. 
In particular, in the small-$x$ region $x < 10^{-3}$, the fit favors a flat form of $S_{Pb}(x) \approx 0.6$ with a
$5$\% accuracy 
at $x=6 \times 10^{-4} - 10^{-3} $ and a $25$\% uncertainty at $x=10^{-4}$.
At the same time, within significant uncertainties, a slow decrease of $S_{Pb}(x)$ with a decrease of $x$ is not ruled out.
At large $x$, $S_{Pb}(x)$ is constrained to better than 10\% accuracy up to $x=0.04$ and is
also consistent with the value of $S_{Pb}(x=0.042)=0.90 \pm 0.10$, which we found from the $A$ dependence 
of the cross section of coherent $J/\psi$ photoproduction
on fixed nuclear targets measured at Fermilab. 
The uncertainties on $S_{Pb}(x)$ are small, which demonstrates the potential of 
the LHC data on coherent charmonium photoproduction in Pb-Pb UPCs to provide new constraints on small-$x$ nPDFs and possible 
signs of saturation. We explicitly demonstrated this using the Bayesian reweighting of the discussed UPC data, which resulted in
a severalfold reduction of the uncertainties of the EPPS16 and nCTEQ15 nPDFs. Also, while the central value of the EPPS16 gluon distribution was not affected by the reweighting, the inclusion of the UPC data favors larger values of the small-$x$ gluon distribution in the nCTEQ15 case.  
It will also be very beneficial to collect high statistics on $J/\psi$ photoproduction in heavy-ion UPCs at RHIC, which would
cover $x \sim 0.015$ at $y=0$ and help to further constrain the results of our analysis, provided that the experimental accuracy is
sufficiently high.

\end{document}